\documentclass[prc,preprint]{revtex4}
\usepackage{bm}
\usepackage{amsmath}
\tighten

\begin{document}
\bibliographystyle{revtex}
\title{Momentum distribution in heavy deformed nuclei: role of
 effective  mass}

\vspace{0.5cm}
\author{
V.O. Nesterenko$^{1,2}$, \fbox{V.P. Likhachev$^2$}, P.-G. Reinhard$^3$,
V.V. Pashkevich$^1$, W. Kleinig$^{1,4}$, and  J. Mesa$^2$
}
\affiliation{ $^{1}$ Bogoliubov Laboratory of Theoretical Physics,
Joint Institute for Nuclear Research 141980, Dubna, Moscow Region, Russia,
E-mail: nester@thsun1.jinr.ru
\\
$^2$ Instituto de F\'{i}sica, Universidade de S\~{a}o Paulo,
S\~{a}o Paulo, E-mail: joel@axpfep1.if.usp.br
\\
$^3$ Institut f\"ur Theoretische Physik,Universit\"at Erlangen,
W-8520 Erlangen, Germany,
E-mail: mpt218@theorie2.physik.uni-erlangen.de
\\
$^4$ Technische Universit\"at Dresden,
Institut f\"ur Analysis, Dresden, D-01062, Germany,
E-mail: kleinig@thsun1.jinr.ru},

\date{\today}

\begin{abstract}
The impact of nuclear deformation on the momentum distributions (MD)
of occupied proton states in $^{238}$U is studied with a
phenomenological Woods-Saxon (WS) shell model and the self-consistent
Skyrme-Hartree-Fock (SHF) scheme.  Four Skyrme parameterizations (SkT6,
SkM*, SLy6, SkI3) with different effective masses are used. The
calculations reveal significant deformation effects in the
low-momentum domain of $K^{\pi}=1/2^{\pm}$ states, mainly of
those lying near the Fermi surface. For other states, the deformation
effect on MD is rather small and may be neglected. The most remarkable
result is that the very different Skyrme parameterizations and the WS
potential give about identical MD.  This means that the value of
effective mass, being crucial for the description of the spectra, is
not important for the spatial shape of the wave functions and thus for
the MD. In general, it seems that, for the description of MD at $0\le k
\le 300$ MeV/c, one may use any single-particle scheme
(phenomenological or self-consistent) fitted properly to the
global ground state properties.
\end{abstract}

\pacs{21.10.Ft, 21.30.Fe, 21.60.-n, 27.90.+b}
\maketitle

\section{Introduction}

The momentum distribution (MD) of nucleons in nuclei is a basic
observable carrying important information on the single-particle
aspects of nuclear structure (see \cite{Ant} and references
therein). In spite of intense studies, there remain several open
points which deserve closer inspection as, e.g., the impact of
nuclear deformation on the MD. The deformation mixes different
spherical components in single-particle wave functions \cite{Gar}
and thus can affect the MD.  The question is how strong the deformation
impact is and what pattern it produces.

In light and rare-earth nuclei, the deformation effects in MD
were earlier studied with the Nilsson model and the Skyrme-Hartree-Fock
(SHF) approach with SIII forces \cite{guer_nil}. Therein, it was also
discussed how the MD signal can be extracted from $(e,e'p)$ reaction.
In the meantime, both experiment and
theory have made substantial progress.  Modern $(e,e'p)$ experiments start to
deal with heavy (actinide) deformed nuclei (see, e.g. \cite{Likh}). At the
theoretical side, the self-consistent SHF approach came into the focus of MD
studies \cite{Ant}. Thus a state-of-art theoretical analysis has to cover
heavy nuclei and include the comparison between a variety of phenomenological
and self-consistent MD descriptions.

In a recent letter \cite{jpg}, we have presented first results on MD in
$^{238}$U by using the WS model and SHF with the SkM* force.  The letter
focused on bound $K^{\pi}=1/2^+$ states with strong $l=0$ contributions at
zero momentum. It was shown that a finite deformation can enhance the number
of such states. These states can be discriminated in knock-out reactions.
Their observation in heavy deformed nuclei can give a valuable information
on the underlying mean field, which, in turn, may be very helpful for
clarifying some hot problems in nuclear structure (see, e.g., \cite{afa}).
The brief study in \cite{jpg} calls for a deeper analysis.  First of all, the
role of the effective mass in SHF models has to be clarified. The effective
mass is known to have a dramatic effect on single-particle spectra
\cite{Mahaux,Brown} but its influence on MD of individual states
is still unclear.  It is thus the aim of the this paper to investigate
in detail the influence of the effective mass on the MD. We will
also give a detailed analysis of general deformation
effects in momentum distributions.

As test case, we will again consider the axially deformed nucleus $^{238}$U.
It is a typical actinide nucleus with well known spectroscopic
characteristics. Actinides are most suitable for our aims.  Indeed, the
heavier is the nucleus, the denser is its single-particle spectrum, and
thus the better are conditions for the deformation mixing.  So, actinides
promise the most strong deformation effects.

Two essentially different single-particle models will be used:
i) the phenomenological Woods-Saxon (WS) potential \cite{Pash}
and ii) the self-consistent density-dependent
Skyrme-Hartree-Fock (SHF) potential \cite{Skyrme}
with Skyrme parameterizations
SkT6 \cite{skt6}, SkM* \cite{skms}, SLy6 \cite{sly6}
and SkI3 \cite{ski3}. These parameterizations represent
different kinds of Skyrme forces and, what is important,
cover a wide interval of the effective masses, from
$m^*/m=1.00$ in SkT6 to  $m^*/m=0.58$ in SkI3
(for an extensive review of self-consistent nuclear models
see \cite{Ben03a}).
The smaller is the effective mass, the more stretched
is the single-particle spectrum (see, e.g. discussions
in Refs. \cite{Mahaux,Brown}) and so the weaker
deformation mixing of spherical configurations is expected.

\section{Formal framework}
\label{sec:model}

The calculations have been performed with the
phenomenological WS and self-consistent SHF potentials.

In case of the WS potential, the deformed shape was described by the
Cassini ovaloids  and the potential-energy surface was calculated as
a function of the elongation $\epsilon$ and hexadecapole deformation
$\alpha_4$ \cite{Pash}. The equilibrium (ground state) deformation
was found by minimizing the total energy. We used the standard set
of the WS  parameters \cite{Chep68} slightly modified for actinides
\cite{WS99}. The single-particle wave functions were expanded in
the  Nilsson basis involving 21 shells.

The SHF calculations were performed with a code using coordinate-space
representation with cylindrical coordinates \cite{skyax}. Four
different Skyrme parameterizations were used: SkT6 \cite{skt6}, SkM*
\cite{skms}, SLy6 \cite{sly6}, and SkI3 \cite{ski3}.
Although these four forces are fitted with different bias, they all
provide a good overall description of nuclear bulk properties and
are equally suitable for heavy deformed nuclei. The essential
feature for our study is that these four forces cover
different values of the effective mass $m^*/m$ (see
Table~\ref{tab:skyrme}).

The bare G matrix theory results in $m^*/m=0.7$ \cite{Bern}.  The same
value is obtained from empirical data for the levels far beyond the
Fermi energy, $|E-E_F|\ge 20$ MeV, in \cite{Mahaux}.  The effective
masses $m^*/m <1$ are known to stretch the single-particle spectra
\cite{Mahaux,Brown}, making them dilute as compared to the experimental
data.  After taking into account the correlation effects, the spectra
should be more compressed and come
closer to the experimental level density.  Actual Skyrme
parameterizations are developed with $m^*/m$ as fitting parameter
varied in a reasonable interval $0.6< m^*/m <1$. The concrete
value for $ m^*/m$ depends on the preferences of observables in the
fit. The giant quadrupole resonance is best fitted with $m^*/m\approx
0.8$, the value for SkM*.  Nuclear surface properties seem to drive to
the lower values found in SLy6 and SkI3. A bias on nuclear energies
complies well with $m^*/m=1$.  It is to be noted that WS and other
phenomenological potentials (Nilsson, etc) employ a "trivial" kinetic
energy, i.e. $m^*/m=1$. Thus the Skyrme
forces with $m^*/m\sim 1$ should in general give spectra close to the
phenomenological ones.

The pairing was treated in the BCS approximation.
The SHF forces used a zero-range two-body pairing force with
strengths adjusted for each parameterization separately, for details
see \cite{Ben03a}. Details of the BCS procedure in the WS case
are given in \cite{Pash}.

\begin{table}
\caption{\label{tab:skyrme}
Effective masses ($m^*/m$) for Skyrme forces, quadrupole
moments ($Q_2$), Fermi energies ($E_F$) and energies of the
lowest ($E_0$) proton single-particle levels in $^{238}$U.
Experimental estimations for the quadrupole moment in
$^{238}$U lie in the interval $Q_2=11.1 - 11.3 $b
\protect\cite{NDS,Sol}.
}
\begin{tabular}{|c|c|c|c|c|}
\hline
Potential & $m^*/m$  & $Q_2$[b] & $E_F$[MeV] & $E_0$[MeV] \\
\hline
 WS    &   -   & 11.66 & -6.63 & -33.69 \\
\hline
 SkT6  & 1.00  & 11.10 & -6.48 & -32.75 \\
 SkM*  & 0.79  & 11.11 & -6.17 & -39.80  \\
 SLy6  & 0.69  & 11.06 & -7.25 & -43.12  \\
 SkI3  & 0.58  & 10.89 & -7.19 & -48.53  \\
\hline
\end{tabular}
\end{table}
Basic ground state properties for the different models are shown in
Table~\ref{tab:skyrme}. The WS potential and all SHF parameterizations
give a reasonable quadrupole moment and Fermi energy for $^{238}$U. At
the same time, they yield different spectral stretching (defined as
a difference, $|E_F-E_0|$, between the Fermi energy (chemical
potential) and the energy of the lowest single-particle level). The
stretching ranges
from 26 to 41 MeV and, as expected,  grows with decreasing the effective mass
(see also Figs. 1 and 2
below). SkT6
with  $m^*/m =1$ gives an average spectral density close to the WS one.

Let's now outline the calculation of MD. The density for the proton
single-particle state $\alpha$   is determined by the standard way
\begin{equation}
 \rho_{\alpha} ({\bf r})=
   |\psi_{\alpha\sigma}({\bf r})|^2  =
   \sum_{\sigma =\pm 1}
  |R_{\alpha}^{(\sigma)}(r,z)|^2
\end{equation}
where
\begin{equation}
\psi_{\alpha}({\bf r})= \sum_{\sigma =\pm 1}
R_{\alpha}^{(\sigma)}(r,z) e^{im_{\alpha}^{(\sigma )}\phi }
\chi_{\sigma}
\end{equation}
is the single-particle wave function of the state
$\alpha $, written in cylindrical coordinates $(r,z,\phi )$.  The
label $\alpha =K^{\pi}[Nn_z\Lambda]$ is composed from the exact quantum
numbers $K^{\pi}$ (total angular momentum projection onto the axial
symmetry axis and parity) and the asymptotic Nilsson quantum numbers
$[Nn_z\Lambda]$.  Further, $\sigma $ is the spin and
$m^{(\pm)}=K\mp 1/2$ is the orbital momentum projection.

In the momentum space $(k_r, k_z, k_{\phi})$, the density
for the state $\alpha$ is defined as
\begin{equation}
\label{eq:rho_k}
  n_{\alpha}({\bf k})=
   |\psi_{\alpha}({\bf k})|^2=
  \sum_{\sigma =\pm 1} |\tilde{R}_{\alpha}^{(\sigma )}(k_r,k_z)|^2
\end{equation}
%
where $\psi_{\alpha\sigma}({\bf k})$ is the Fourier-transformed
single-particle wave function. In the WS potential, the wave functions
are expanded in the Nilsson basis whose Fourier-transformation is
done analytically (for more details see,
e.g. \cite{guer_nil}).  In SHF, the Fourier-transformed wave function
reads as
\begin{eqnarray}
\psi_{\alpha}({\bf k})&=&\frac{1}{(2\pi )^{3/2}}
\int^{\infty}_{-\infty}dz \int_0^{\infty} r dr\int_0^{2\pi}
d\phi
\nonumber\\
&\cdot & e^{i{\bf kr}} \sum_{\sigma =\pm 1} R_{\alpha}^{(\sigma )}(r,z)
e^{im_{\alpha}^{(\sigma )}\phi}
\nonumber \\
&=&
\sum_{\sigma =\pm 1} \tilde{R}_{\alpha}^{(\sigma )}(k_r,k_z)
e^{im_{\alpha}^{(\sigma )}k_{\phi}} i^{m_{\alpha}^{(\sigma )}}
\end{eqnarray}
where
\begin{equation}
\tilde{R}_{\alpha}^{(\sigma )}(k_r,k_z) =
\frac{1}{\sqrt{2\pi}}\int^{\infty}_{-\infty} dz \int_0^{\infty}
dr R_{\alpha}^{(\sigma )}(r,z) j_{m_{\alpha}^{(\sigma )}}(k_rr)
e^{ik_zz}
\end{equation}
and $j_{m_{\alpha}^{(\sigma )}}(k_rr)$ is the Bessel function.

As is usually done in $(e,e'p)$ calculations, we average
(\ref{eq:rho_k}) over the nuclear symmetry axis direction:
\begin{equation}
\label{eq:rho_av}
  n_{\alpha}(k)=\frac{1}{2} \int_0^{\pi}
d\theta \sin\theta  n_{\alpha}({k_r,k_z})
\end{equation}
where $k_r=k\sin\theta$ and $k_z=k\cos\theta$.
%

It is
worth noting that in general single-particle models are not well
suited to describe MD because of the important contributions from
short- and long-range correlations \cite{Ant}.  However these
perturbing effects take place mainly in the high-momentum domain with
$k > k_{\rm F}\approx 1.3 \; {\rm fm}^{-1} \approx 260\,{\rm MeV/c}$
while we will focus on MD at low $k$, where the single-particle models
are still appropriate.

\section{Results and discussion}
\subsection{Single-particle levels}

Before discussing the MD for single-particle states, it is instructive
to have a look at the single-particle spectra.  Figs. 1 and 2
compare WS and SHF single-particle proton and neutron spectra in $^{238}$U.
To make the analysis more transparent, the level schemes are
presented in the spherical limit. This avoids the complexity
caused by the deformation splitting of the levels and thus
concentrates on the essential trends. The figures demonstrate the
stretching of the single-particle spectra with decreasing the effective
mass $m^*/m$. While the Fermi energies remain basically at
the similar position, the hole levels steadily dive deeper from SkT6 with
$m^*/m=1.00$ to SkI3 with $m^*/m=0.58$. In agreement with
Ref. \cite{Mahaux}, the main stretching effect takes place for the
spectra far from the Fermi energy, first of all for the deeply bound
levels. The spectra near the Fermi energy also show changes but not so
strong and regular, thus mainly displaying the influence of other
Skyrme terms, in particular of the spin-orbit force. As is expected,
the SkT6 spectra with $m^*/m=1.00$ are most similar to the WS
ones. This is most obvious for the case of neutrons. The proton WS spectrum
looks somewhat wider, which can be partly explained by the downshift
of the Fermi level in the WS case. In any case, this difference is not
important. It is more essential that the WS and SkT6 demonstrate
very similar {\it global} spans between the Fermi and the lowest levels (see
Table \ref{tab:skyrme}).


The stretching of the spectra with decreasing $m^*/m$ can be
understood if one assumes that the system keeps the average momenta
$k_{\alpha}$ for single-particle states. As is shown below, this is
indeed the case. Then the kinetic energies
$T_\alpha= k^2_{\alpha}/(2m^*)$
increase with decreasing $m^*$, and so, to keep the same Fermi
energy, the depth of the potential needs to be increased as
well. Since the kinetic energy for deep hole states is smaller than for
the valence states, the relative kinetic (smaller) and potential
(larger) shifts result in pushing the lowest states deeper down and
thus in stretching the spectrum.

Figure 3
shows the proton levels for the deformed ground
state in the vicinity of the Fermi surface. We see again that the
Skyrme spectra become in general more dilute with decreasing
$m^*/m$. However, the trend is much weaker than for deeply lying states
and concerns only the uppermost and lowest states in the plot.  The
intermediate states do not exhibit any clear relation to $m^*/m$,
like in the spherical case.
The BCS calculations of the quasiparticle spectra reveal
that only the WS and SkM* reproduce the correct
ground states assignment of the neighboring odd-proton nuclei
$^{239}$Np ($5/2^+$) and $^{237}$Pa ($1/2^+$).  This result, however,
does not allow to judge on the accuracy of the single-particle schemes.
As was mentioned above, the comparison with experiment for the odd nuclei
requires, in principle, the inclusion of polarization effects
\cite{Sol,Rutz_ug}, in particular the coupling with
core vibrations.

In that connection, it is worth noting that SHF is not always
performing well with the spectra of particular
nuclei (see, e.g.  discussion in Ref. \cite{Brown}). This can be
explained (and excused) by the fact that most of the Skyrme
parameterizations are tuned to optimize the basic ground state
characteristics (binding energies, r.m.s. charge radii, densities, etc)
instead of the spectra. Moreover, recent Skyrme forces aim to
describe nuclei, both spherical and deformed, throughout the entire mass
table (including those near drip-lines) as well as nuclear and neutron
matters. Certainly, such universality has a price:
Skyrme forces are generally not optimized to deliver optimal spectra
(unlike phenomenological potentials which are usually specially fitted
to low-energy spectra in particular nuclei).

\subsection{Momentum distributions}

%
%

In Fig. 4,
the MD from SkM* calculated at the equilibrium
shape and in the spherical limit are compared for a representative set
of nine occupied proton states.  Both deeply and slightly bound states
are involved.  The deep hole states include $1/2^-[330]$ and
$1/2^-[301]$ (with the single-particle energies -26.5 and -17.2 MeV,
respectively). The other seven states lie near the Fermi energy (see
Fig. 3.)
As is discussed below, the weakly bound states
are expected to deliver the most pronounced deformation effects and
thus we pay to them more attention.  In the spherical limit, the
number of maxima in the MD profile for the state $nlj$ is equal to the
number of radial nodes $n$. Hence, the deformation effects can be
easily spotted by looking at an increasing number of the maxima and/or
an essential redistribution of the strength between the maxima. We
present here the SkM* results though, as is discussed below, other Skyrme
parameterization might be used as well.

Figure 4
illustrates some general deformation effects in
MD.  First, the lower the $K$ quantum number, the stronger the
deformation effect, see e.g. the $K=1/2$ states $1/2^-[330]$ and
$1/2^-[530]$ (and the state $1/2^+[660]$ in Fig. 5).
This
follows from the fact that spherical configurations with low $K$ have
in general a denser spectrum than those with high $K$, which favors
the mixing low-$K$ states due to the deformation. The exceptions
(e.g., $1/2^-[301]$) mainly concern deeply bound states whose mixing
is often suppressed due to a rather dilute spectrum. At the other
side, the levels near the Fermi energy are more affected by the
deformation because they reside in a region of higher spectral
density.

Second, the deformation usually results in a shift of the MD strength
to lower momenta.  Note that the normalization condition $\int n_{\alpha}(k)
k^2 dk=1$ carries a weight $k^2$ and so even a small modification of
MD at high $k$ may cause considerable changes at low $k$. As a result,
just the low-$k$ domain is most sensitive to deformation (see also the
discussion on $K^{\pi}=1/2^+$ states in \cite{jpg}).

Figure 5
compares MD from the WS and SkM*.
We see rather good agreement between both cases. The modest differences
mainly take place in the low momentum
regions where deformation effects are most strong. This result somewhat
deviates from that in \cite{jpg} where the deformation effects in the
WS potential were overestimated because of the insufficiently accurate
treating of the WS wave functions in the momentum space.

It worth noting that the calculations \cite{guer_nil} for Ne and Nd
also displayed rather modest differences between
phenomenological (Nilsson) and Skyrme (SIII) momentum distributions
and the deformation effect mainly in the low-momentum domain
of the $K=1/2$  states. At the same time, our calculations predict
more cases of the noticeable deformation impact since we deal
with the heavier nucleus  $^{238}$U where the deformation
mixing is generally stronger.

Altogether, one may conclude that just the $K=1/2$ states in heavy
nuclei, lying in the vicinity of the Fermi energy, are most promising
for displaying the deformation effect in MD. In deeply bound states,
even if they are influenced by the deformation, the momentum
distributions should be considerably smeared by polarization as well
as correlations, thus hiding, to a large extend, the deformation
mixing. The only chance for deeply bound states to exhibit in experiment
the deformation effects is offered by the $K^{\pi}=1/2^+$
states, where the deformation induced $l=0$ strength is strictly
localized at $k=0$ and so can in principle be distinguished from the
$l \ne 0$ patterns \cite{jpg}.

Figure 6
compares MD for the four different Skyrme forces
(SkT6, SkM*, SLy6, and SkI3). In spite of the much
different effective masses, all the parameterizations give very similar
MD.  The deviations are about invisible at high momenta for all the
states and at all momenta for the states with high $K$.  Even for deep
hole states $1/2^-[301]$ and  $1/2^-[330]$, the MD are about the same.
The minor
deviations related to the deformation are spotted in $1/2^-[330]$ and
$1/2^-[530]$. The only strong difference takes place at low $k$ in
$1/2^+[660]$. However, this case is very specific and reflects the
considerable mixing of $1/2^+[660]$ and $1/2^+[400]$ states, which is
well known in deformed nuclei.  Just because of the $1/2^+[400]$ admixture
with its dominant $3s_{1/2}$ component, the state $1/2^+[660]$
acquires a jump at $k=0$.  The mixing $1/2^+[660]$ and $1/2^+[400]$ levels
is caused by their pseudo-crossing at the equilibrium deformation.
The states at the crossing points are known to be
extremely sensitive to the details of the single-particle scheme and
in this sense the state $1/2^+[660]$  is an
exception from the general picture. So, we may conclude that the value of
the effective mass, being crucial for the description of the spectra,
turns out to be irrelevant for the momentum distributions.

The insensitivity of MD to the effective mass is the most remarkable
result of our study. It means that single-particle wave
functions are much less sensitive to the effective mass than the
single-particle spectra. Moreover, the similarity of MD obtained with
different potentials (Nilsson, WS, SHF) signifies that independence
of MD to the effective mass is a signature of a more general robustness
of MD. In principle, this feature is not surprising since
MD are determined by the structure of the wave functions which in turn
is specified by the orbital moment and number of nodes of the dominant
components. All the relevant single-particle potentials evidently keep
this structure in the spherical limit. In deformed nuclei, the
different models should reproduce the nuclear quadrupole moment and
then their eigenfunctions should have a similar composition of
angular momentum components which yields, in turn, similar MD.  We
thus may conclude that any single-particle potential (phenomenological or
self-consistent) which reproduces the basic ground
state properties should accurately describe momentum distributions of
individual states in the momentum domain $0\le k \le 300$ MeV/c
(for the exception of the cases of the level crossing).

General arguments given above are still not enough for treating so
nontrivial result as the indifference of the SHF MD to the effective mass
and we need here some additional comments. It would be natural to
expect the similarity of MD for the different phenomenological
potentials which deviate only by the potential term while the kinetic
term remains to be the same. But, in SHF forces with various $m^*$,
both the potential and kinetic parts are different. And, if we get
then very different SHF single-particle spectra, why not
to expect also the different MD, at least for the deep hole states?
Indeed, the SHF parameterizations are
fitted so as to reproduce the ground state properties which are mainly
determined by the nucleons from the valence shell. But deep hole
states should not be so fixed by the fit and so might in principle
deviate not only in spectra but also in MD. For example, the  MD peaks
might be somewhat shifted. However, MD of the deep hole states persist
to keep their profiles at different $m^*$. It looks like
the valence shell strictly determines $n_{\alpha}(k)$ and
$n_{\alpha}(r)$ (and thus the single-particle wave functions) in other
shells as well, and the nucleus preserves the velocity fields for all
the nucleons, both valence and deep hole.

\section{Conclusions}

The influence of the nuclear deformation on the momentum distributions
(MD) of proton hole states in $^{238}$U was studied with the phenomenological
WS potential and the self-consistent Skyrme-Hartree-Fock approach.
Four Skyrme parameterizations (SkT6, SkM*, SLy6, and SkI3) with
effective masses $ 0.58\le m^*/m \le 1.00$ were used.  Particular
attention was paid to the role of the effective mass.

It was shown that the main deformation effects take place at the
low-momentum domain of $K^{\pi}=1/2^{\pm}$ states in the vicinity of
the Fermi energy.
Indeed, the lower angular momentum projection $K$ have the states, the
larger is their average spectral density. Besides, the spectral
density rises with approaching the Fermi energy. The high
spectral density favors the deformation mixing.  As a result,
just $K^{\pi}=1/2^{\pm}$ levels near the Fermi energy are mainly
affected by the deformation.

The most striking result concerns the role of the effective mass. The
calculations confirm that the effective mass strongly influences the
single-particle spectrum. At the same time, different Skyrme forces
with the effective masses varying trough $0.58\le m^*/m \le 1$, give
about identical MD. The remaining modest differences in MD are mainly
connected with the deformation effects. Such a striking similarity of
Skyrme MD leads to the surprising (at the first glance) conclusion that
the effective mass does not influence the momentum distributions in
a nucleus. The deviations in other features of the Skyrme forces also do
not noticeably
influence MD. Moreover, the Skyrme MD are quite similar to the
WS ones. So, for the description of MD (and the subsequent inputs for
knock-out reactions) one can use any well fitted Skyrme or
phenomenological potentials. We mainly explain such stability of the momentum
distributions as a consequence of the fact that any single-particle
potentials properly fitted to the basic ground state properties
(including the nuclear shape) keep the same structure (principle
components with their orbital moments and node numbers) of the wave
functions.

It is interesting that, though mainly the valence shell
is responsible for the ground state properties, the
momentum distributions of the deep hole states also become fixed by
the fit. Unlike the phenomenological potentials, SHF forces with
different $m^*$ deviate not only in the potential term but in
the kinetic energy as well. And this is a nontrivial and somewhat
unexpected result that MD, unlike the
spectra, turn out to be so stable even for the deep hole states.

\begin{acknowledgments}
  We thank Profs. A.N. Antonov and S. Frauendorf for fruitful discussions.
  This work was supported by the grants  FAPESP No. 2001/06082-1,
Germany-BLTP JINR (Heisenberg-Landau) and
Bun\-des\-ministerium f\"ur Bildung und Forschung (BMBF),
Project No.\ 06 ER 808.
\end{acknowledgments}


\begin{thebibliography}{99}
\bibitem{Ant} 
  A.N. Antonov, P.E. Hodgson, and I.Zh. Petkov,
 {\it Nucleon Correlations in Nuclei}
 (Springer-Verlag Berlin Heidelberg, 1993).
\bibitem{Gar}
  F.A. Gareev, S.P. Ivanova, L.A. Malov, and V.G. Soloviev,
  Nucl. Phys. A{\bf 171}, 134 (1971).
\bibitem{guer_nil} 
 J.A. Caballero and E. Moya de Guerra,
 Nucl. Phys. A{\bf 509}, 117 (1990);
 E. Moya de Guerra, P. Sarriguren, J.A. Caballero, M. Casas,
 and D.W.L. Sprung, Nucl. Phys. A{\bf 529}, 68 (1991).
\bibitem{Likh}
  V.P. Likhachev, J. Mesa, J.D.T. Arruda-Neto,
  B.V. Carlson, A. Deppman, M.S. Hussein, V.O. Nesterenko,
  F. Garcia, and O. Rodriguez,
  Phys. Rev. C{\bf 65}, 044611 (2002).
\bibitem{jpg}
    V.O. Nesterenko, V.P. Likhachev, P.-G. Reinhard,
    J. Mesa, W. Kleinig, J.D.T. Arruda-Neto, and A. Deppman,
    J. Phys. G: Nucl. Part. Phys. {\bf 29}, L37 (2003).
\bibitem{afa}
  A.V. Afanasjev, T.L. Khoo, S. Frauendorf, G.A. Lalazissis, and
  I. Ahmad,  Phys. Rev. C{\bf 67}, 024309 (2003).
\bibitem{Mahaux}
  C. Mahaux, P.F. Bortignon, R.A. Broglia, and C.H. Dasso,
  Phys. Rep. {\bf 120}, 1 (1985).
\bibitem{Brown}
  B.A. Brown,
   Phys. Rev. C{\bf 58}, 220 (1998).
\bibitem{Pash}
   V.V. Pashkevich,
   Nucl. Phys. A{\bf 169}, 275 (1971).
\bibitem{Skyrme}
  T.H.R. Skyrme,
  Phil. Mag. {\bf 1}, 1043 (1956);
  D. Vauterin and D.M. Brink,
  Phys. Rev. C{\bf 5}, 626 (1972).
\bibitem{skt6}
  F. Tondeur, M. Brack, M. Farine, and J.M. Pearson,
  Nucl. Phys. A{\bf 420}, 297 (1984).
\bibitem{skms}
  J. Bartel, P. Quentin, M. Brack, C. Guet, and
  H.-B. H\aa{a}kansson,
  Nucl. Phys. A{\bf 386}, 79 (1982).
\bibitem{sly6}
  E. Chabanat, P. Bonche, P. Haensel, J. Meyer, and
  R. Schaeffer,
  Nucl. Phys. A{\bf 643}, 441(E) (1998).
\bibitem{ski3}
  P.-G. Reinhard and H. Flocard,
  Nucl. Phys. A{\bf 584}, 467 (1995).
\bibitem{Ben03a}
  M. Bender, P.-H. Heenen, and P.-G. Reinhard,
   Rev. Mod. Phys. {\bf 75}, 121 (2003)
\bibitem{Chep68}
  V.A. Chepurnov,
  Sov. J. Nucl. Phys. {\bf 6}, 696 (1968).
\bibitem{WS99}
  F. Garcia, E. Garrote, M.-L. Yoneama, J.D.T. Arruda-Neto,
  J. Mesa, F. Bringas, J.F. Dias, V.P. Likhachev,
  O. Rodriguez, and F. Guzm\'{a}n,
  Eur. Phys. J. A{\bf 6}, 49 (1999).
\bibitem{skyax}
  P.-G. Reinhard, unpublished.
\bibitem{NDS}
   E.N. Shurshikov,
   Nucl. Data Sheets, {\bf 53}, 601 (1988).
\bibitem{Sol}
   V.G. Soloviev,
   {\it Theory of Complex Nuclei} (Pergamon Press, 1976).
\bibitem{Bern}
      V. Bernard and N. Van Giai,
      Nucl. Phys. A{\bf 348}, 75 (1980).
\bibitem{Rutz_ug}
  K. Rutz, M. Bender, P.--G. Reinhard, J.A. Maruhn, and W. Greiner,
  Nucl. Phys. A{\bf 634}, 67 (1998)

\newpage
{\bf \large FIGURE CAPTIONS}

\vspace{0.5cm}\indent
{\bf Figure 1}:
WS and SHF proton single-particle spectra
in $^{238}$U at zero deformation (spherical limit).
The effective mass decreases from $m^*/m=1.00$ in SkT6 to
$m^*/m=0.58$ in SkI3.
The levels of the positive and negative parity are depicted
by the solid and dotted lines, respectively. For the view
convenience, the identical levels are connected
by dashed lines. The chemical potentials are indicated
by dotted lines with crosses.

\vspace{0.5cm}\indent
{\bf Figure 2}:
The same as in Fig. 1 for the neutron spectra.

\vspace{0.5cm}\indent
{\bf Figure 3}:
WS and SHF proton single-particle spectra near
the Fermi energy in $^{238}$U. The effective mass decreases from
$m^*/m=1.00$ for SkT6 to $m^*/m=0.58$ for SkI3. The level
energies are given relative to the Fermi level, $1/2^+[400]$
in WS and $3/2^+[651]$ in Skyrme potentials.

\vspace{0.5cm}\indent
{\bf Figure 4}:
Momentum distributions for nine occupied proton states
in $^{238}$U calculated with SkM* in the spherical limit
(solid line) and at the equilibrium deformations (dashed line).
The spherical ancestors are indicated for every state.

\vspace{0.5cm}\indent
{\bf Figure 5}:
SkM* (solid line) and WS (dashed line) proton momentum distributions in
$^{238}$U.

\vspace{0.5cm}\indent
{\bf Figure 6}:
Proton momentum distributions in $^{238}$U calculated with Skyrme potentials
SkT6 (dashed line), SkM* (solid line), SLy6 (dotted line), and SkI3
(dashed-dotted line).


\end{thebibliography}
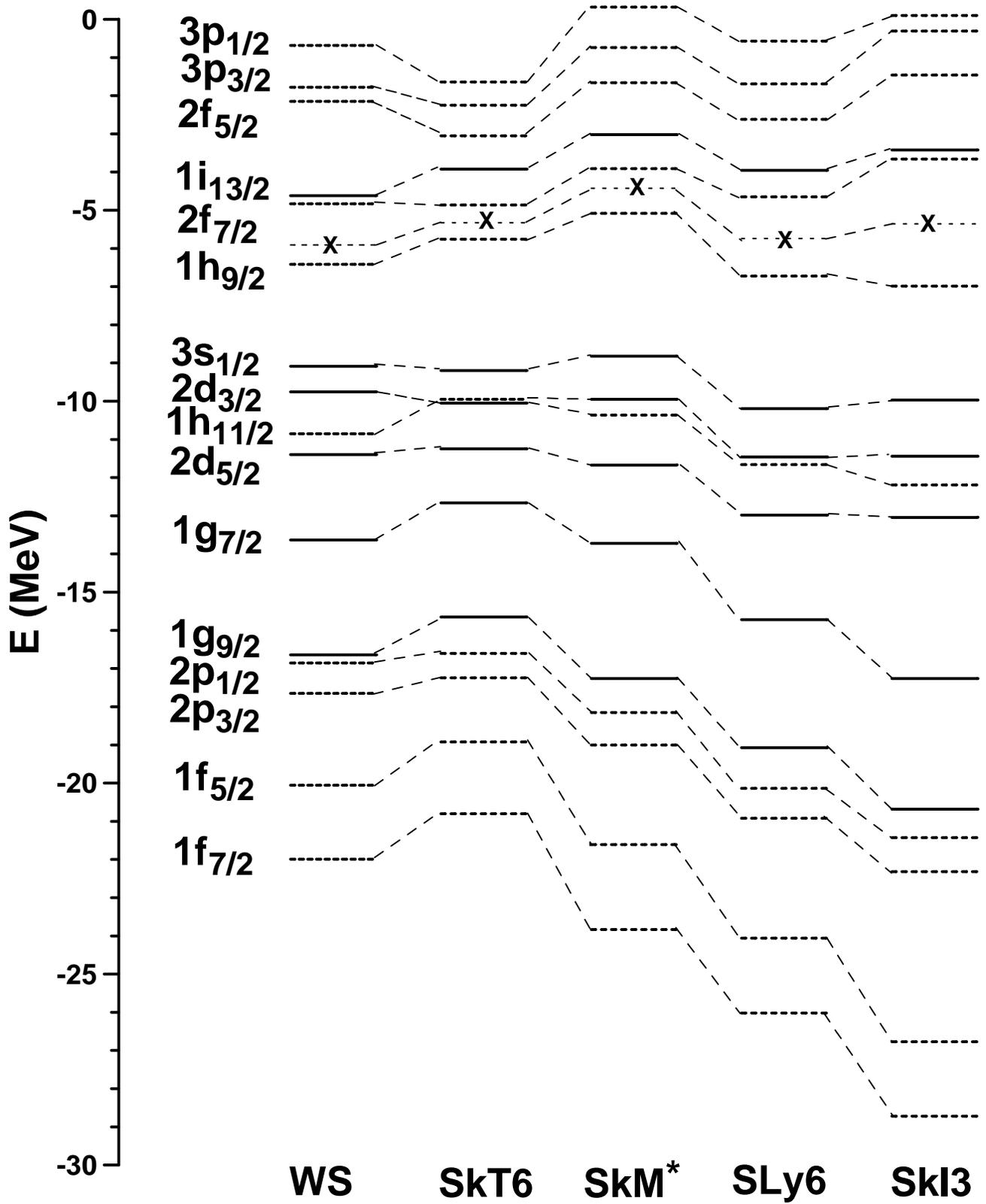
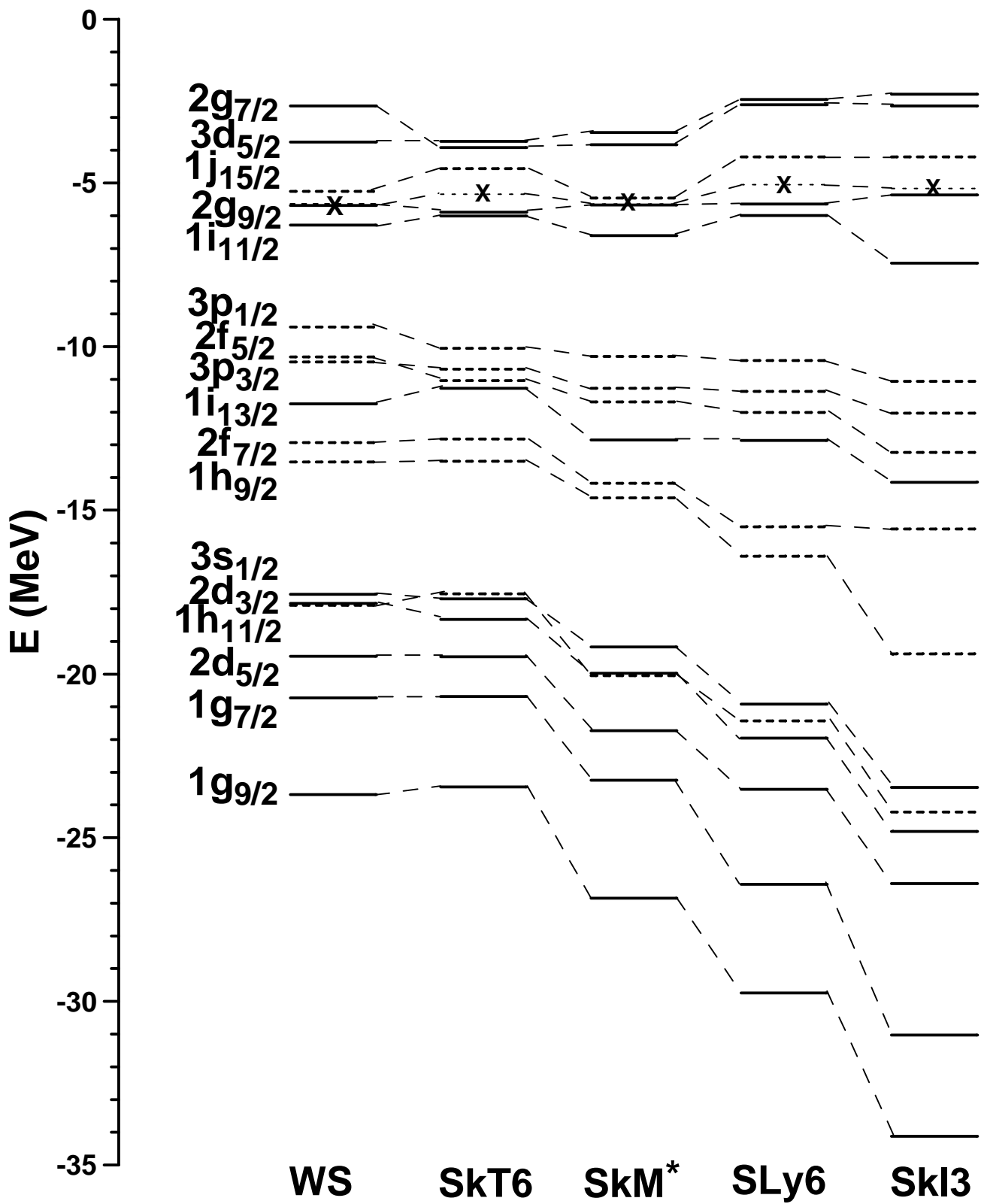
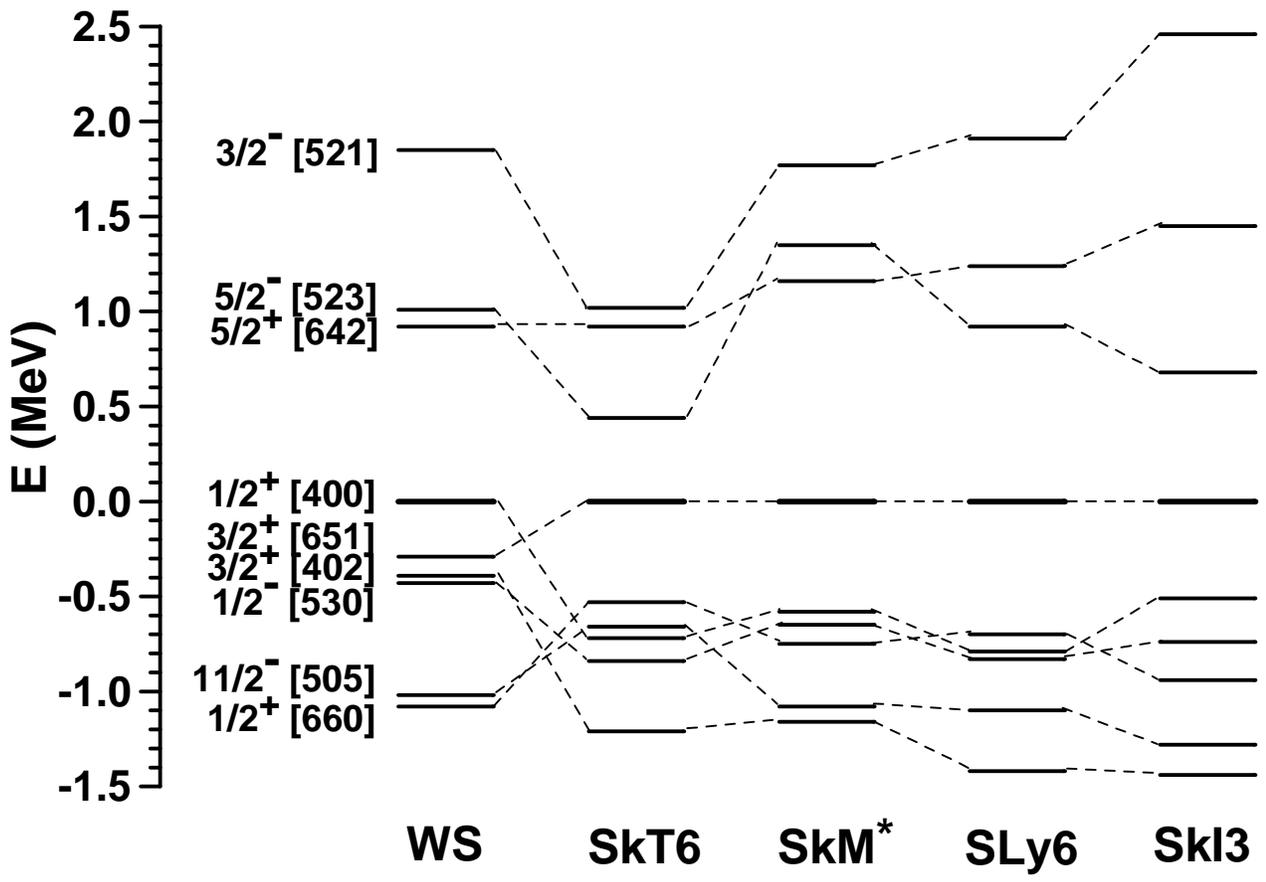
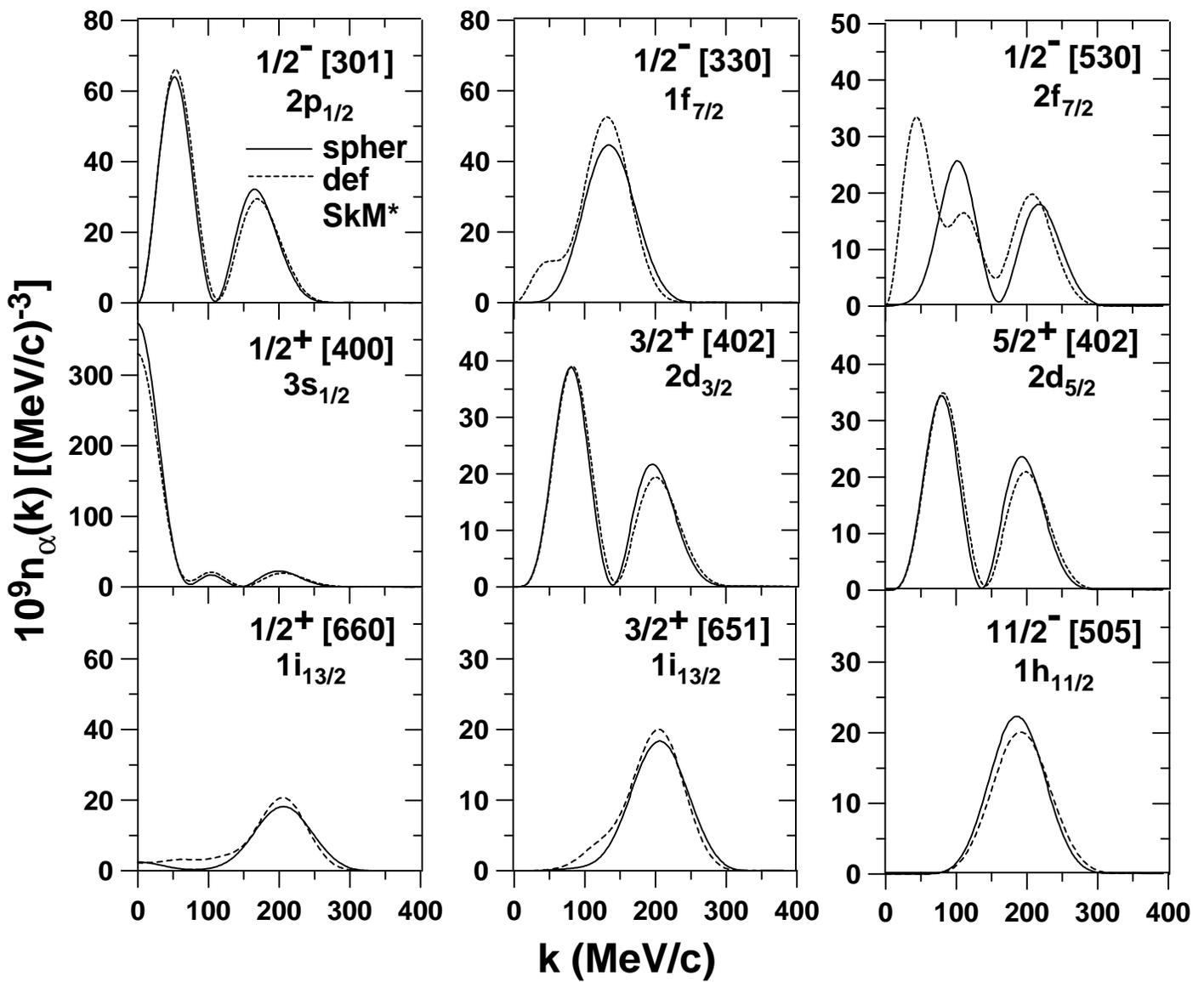
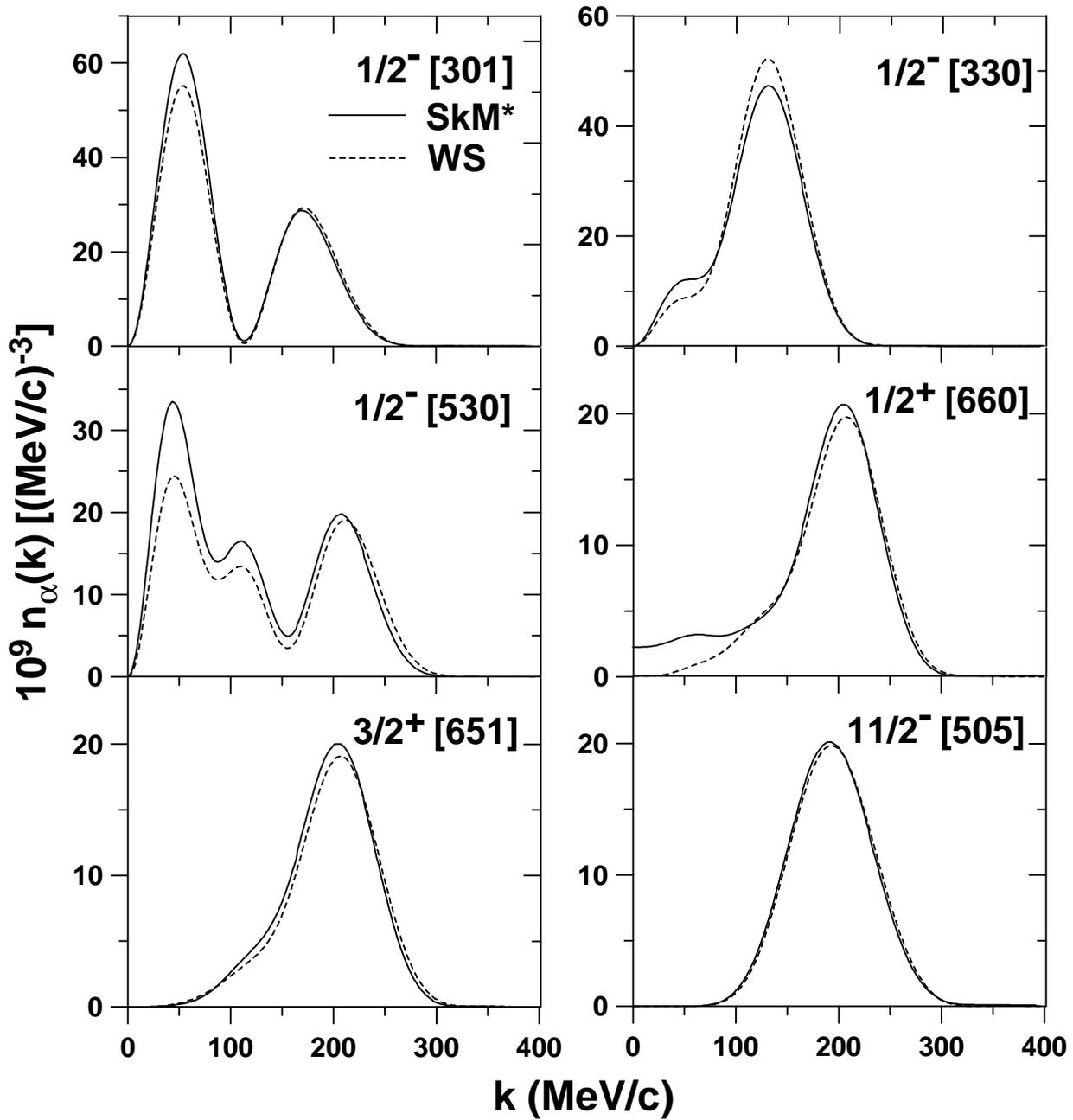
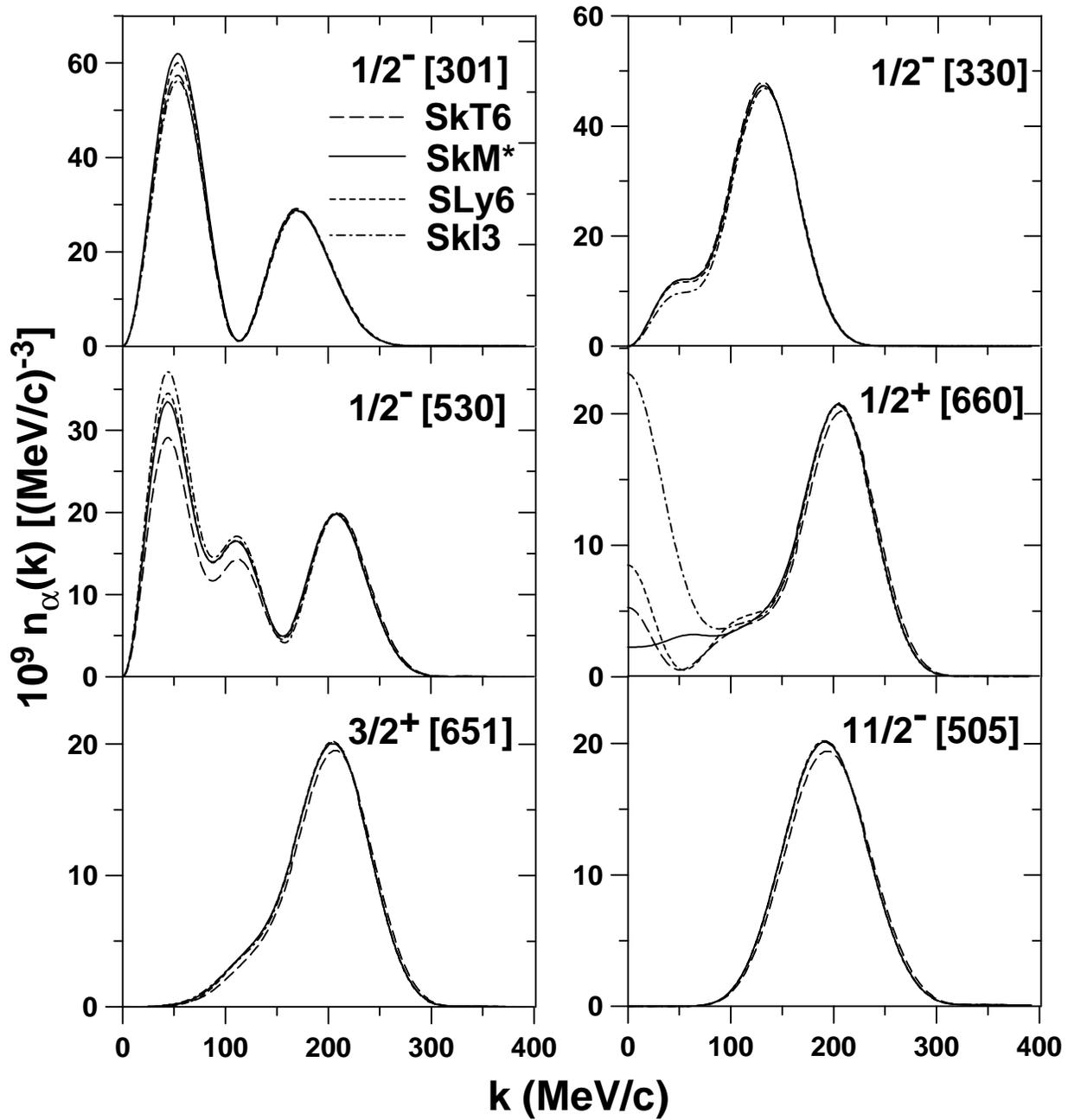
\end{document}